\documentclass[usenatbib]{mn2e}
\usepackage{graphicx}

\title[New southern SU UMa-type dwarf nova, BF Ara]
{Photometric study of new southern SU~UMa-type dwarf novae -- II:
Authentication of BF Ara as a Normal SU~UMa-type Dwarf Nova with the Shortest
Supercycle}

\author[T. Kato et al.]
{\parbox[t]{\textwidth}{
       Taichi Kato$^1$,
       Greg Bolt$^2$,
       Peter Nelson$^3$, 
       Berto Monard$^4$, 
       Rod Stubbings$^5$, \\
       Andrew Pearce$^6$,
       Hitoshi Yamaoka$^7$,
       Tom Richards$^8$ \\
} \\
       $^1$ Department of Astronomy, Faculty of Science,
       Kyoto University, Sakyo-ku, Kyoto 606-8502 Japan \\
       $^2$ 295 Camberwarra Drive, Craigie, Western Australia 6025,
       Australia \\
       $^3$ RMB 2493, Ellinbank 3820, Australia \\
       $^4$ Bronberg Observatory, PO Box 11426, Tiegerpoort 0056,
       South Africa \\
       $^5$ 19 Greenland Drive, Drouin 3818, Victoria, Australia \\
       $^6$ 32 Monash Ave, Nedlands, WA 6009, Australia \\
       $^7$ Faculty of Science, Kyushu University, Fukuoka 810-8560,
       Japan \\
       $^8$ Woodridge Observatory, 8 Diosma Rd, Eltham, Vic 3095, Australia \\
}

\date{Accepted.
      Received;
      in original form}

\pagerange{\pageref{firstpage}--\pageref{lastpage}}
\pubyear{2003}

\begin{document}

\maketitle

\label{firstpage}

\begin{abstract}
   We photometrically observed the 2002 August long outburst of BF Ara.
The observation for the first time unambiguously detected superhumps
(average period 0.08797(1) d), qualifying BF Ara as a genuine SU UMa-type
dwarf nova.  An analysis of the long-term visual light curve yielded a mean
supercycle length of 84.3(3) d.  The characteristics of outbursts and
superhumps more resemble those of usual SU UMa-type dwarf novae rather
than those of ER UMa stars.  BF Ara is thus confirmed to be the usual
SU UMa-type dwarf nova with the shortest known supercycle length.
There still remains an unfilled gap of distributions between ER UMa stars
and usual SU UMa-type dwarf novae.  We detected a zero period change of
the superhumps, which is quite unexpected from our previous knowledge.
This discovery implies that a previous interpretation requiring
a low $\dot{M}$ would be no longer valid, or that a different mechanism is
responsible for BF Ara.  We propose that the reduced (prograde) apsidal
motion of the eccentric disk by pressure forces may be responsible for
the unusual period change in BF Ara.
\end{abstract}


\section{Introduction}

   ER UMa stars are still an enigmatic subgroup of SU UMa-type
dwarf novae (for a review of dwarf novae and SU UMa-type dwarf novae,
see \citet{osa96review} and \citet{war95suuma}, respectively).
Although most of SU UMa-type dwarf novae have supercycle lengths
($T_{\rm s}$: the interval between successive superoutbursts) long than
$\sim$100 d (cf. \citealt{nog97sxlmi}), ER UMa stars have extremely
short $T_{\rm s}$ (19--50 d, for a review, see \citealt{kat99erumareview}).
Only five definite members have been discovered: ER UMa
(\citealt{kat95eruma,rob95eruma,mis95PGCV});
V1159 Ori (\citealt{nog95v1159ori,pat95v1159ori});
RZ LMi (\citealt{rob95eruma,nog95rzlmi});
DI UMa (\citealt{kat96diuma}); and IX Dra (\citealt{ish01ixdra}).

   From the theoretical standpoint, ER UMa stars pose difficult and
interesting problems.  The outburst mechanism of SU UMa-type
dwarf novae is now widely believed to be a combination of thermal and
tidal instabilities in the accretion disk \citep{osa89suuma}.
A smooth extension of SU UMa-type dwarf novae toward higher mass-transfer
rates ($\dot{M}$) seems to be a natural explanation of extremely short
$T_{\rm s}$ in ER UMa stars \citep{osa95eruma}.
This explanation, however, requires a poorly understood mechanism to
prematurely quench superoutbursts to reproduce the extremely short
$T_{\rm s}$ ($\sim$19 d) in RZ LMi \citep{osa95rzlmi}.

   The origin of the supposed high mass-transfer rates is also a mystery,
since the mass-transfer is mainly driven by gravitational wave radiation
in SU UMa-type dwarf novae, within the standard evolutionary framework
of cataclysmic variables (CVs)
(\citealt{rap82CVevolution,rap83CVevolution}; for recent reviews of CV
evolution, see \citealt{kin88binaryevolution,kin00CVevolution}).

   Several attempts have been made to ascribe such a high $\dot{M}$ to
a nova-induced enhancement of mass-transfer (originally discussed in
\citet{nog95rzlmi} in the context of ``nova hibernation'' scenario:
\citealt{Hibernation}).  Recent model calculations, however, have not been
successful to reproduce the supposed wide $\dot{M}$ diversity in
short-period systems to which ER UMa stars belong \citep{kol01mdotvar}.
An irradiation-induced, cyclic mass-transfer variation has been shown
to be also less effective in short-period systems
(\citealt{kin95masstransfercycle,kin96masstransfercycle,mcc98masstransfercycle},
see also a general discussion in \citealt{pat98evolution}).

   Most recently, several ideas have been proposed to explain the unusual
outburst properties of ER UMa stars.  \citet{hel01eruma} proposed an idea to
explain the ER UMa-type phenomenon by considering a decoupling between
the thermal and tidal instabilities.  \citet{bua01DNoutburst}
tried to explain the ER UMa-type phenomenon by introducing an inner
truncation of the accretion disk and irradiation on the secondary star.
These ideas either require a still poorly understood mechanism or
an arbitrary parameter selection, which does not yet seem to reasonably
reproduce observations \citep{bua02suumamodel}.

   From the observational side, the distribution of $T_{\rm s}$ seems to be
discontinuous between ER UMa stars and usual SU UMa-type dwarf novae (c.f.
\citealt{nog97sxlmi,hel01eruma}).  Furthermore, the distribution of the
orbital periods ($P_{\rm orb}$) or superhump periods ($P_{\rm SH}$)
of ER UMa stars strongly concentrates in a short-period region
\citep{ish01ixdra}.  These observational properties have raised the following
central problems: (1) Do ER UMa stars and usual SU UMa stars comprise
a continuous distribution of $T_{\rm s}$? and (2) Are there
long-$P_{\rm orb}$ (or long-$P_{\rm SH}$) ER UMa stars?  These fundamental
questions have not been yet answered.  The second question is particular
important because the working hypotheses by \citet{hel01eruma} and
\citet{bua01DNoutburst} either require a small binary mass-ratio
($q$ = $M_2$/$M_1$) or a short orbital period, which would enable
a weak tidal torque or a strong effect of irradiation, respectively.

   From these motivations, a search for transitional objects between ER UMa
stars and usual SU UMa-type stars, and long-$P_{\rm orb}$ ER UMa stars
has been undertaken.  CI UMa ($T_{\rm s} \sim$ 140 d) was once claimed to be
a transitional object \citep{nog97ciuma}, but the pattern of its
outbursts is much more irregular than those of ER UMa stars. 
A short $T_{\rm s}$ (89 d) system, V503 Cyg \citep{har95v503cyg} is
also unusual in its infrequent normal outbursts.\footnote{
  \citet{kat02v503cyg} reported the detection of a dramatic changes in
  the outburst pattern of V503 Cyg.  V503 Cyg may be a system with two
  distinct states (the states with low or high number of normal outbursts),
  both of which are unlike those of ER UMa stars.
}
Low-amplitude SU UMa-type dwarf nova, HS Vir, which has
similar properties to ER UMa stars in its high frequency of normal
outbursts \citep{kat95hsvir,kat98hsvir}, has recently confirmed to
have a long ($T_{\rm s}$ = 186 or 371 d, \citealt{kat01hsvir}) supercycle,
which is unlike those of ER UMa stars.
Most recently, SS UMi ($T_{\rm s}$ = 84.7 d, \citealt{kat00ssumi})
has been shown to be the shortest $T_{\rm s}$ system having usual
properties of SU UMa-type dwarf novae \citep{kat98ssumi}.  A search for
transitional objects or long-$P_{\rm orb}$ ER UMa has been unsuccessful.

   BF Ara is a dwarf nova having a range of variability 13.6 -- (16.0p
and a tentative classification of an SS Cyg-type dwarf nova according to
the 4-th edition of the General Catalogue of Variable Stars \citep{GCVS}.
\citet{bru83bfara} photometrically studied this object during an outburst,
and recorded 0.25-mag variations which could be attributed to a superhump.
However, because of the lack of a sufficiently long series of photometry
and the lack of knowledge in the outburst properties, this object has
been largely neglected in the past studies.  In the most recent years,
\citet{kat01bfara} noticed the presence of a clear recurring periodicity
of long outbursts (likely superoutbursts).  From an analysis of the
visual observations reported to the VSNET Collaboration,\footnote{
http://www.kusastro.kyoto-u.ac.jp/vsnet/.} \citet{kat01bfara} proposed
a mean supercycle length of 83.4 d, on the presumed assumption that
BF Ara is an SU UMa-type dwarf nova.  Since this supercycle length broke
the shortest record among usual SU UMa-type dwarf novae, BF Ara has been
regarded as a key object to study the borderline and the
relation between usual SU UMa-type dwarf novae and ER UMa stars.
The next important step has undoubtedly been an unambiguous detection of
superhumps which authenticates BF Ara as a genuine SU UMa-type dwarf nova.
We conducted a photometric campaign during a long outburst in 2002
August as an intensive project of the VSNET Collaboration
\citep{kat03v877arakktelpucma}.

\begin{table}
\caption{Observers and Equipment.} \label{tab:equipment}
\begin{center}
\begin{tabular}{cccc}
\hline\hline
Observer   & Telescope &  CCD  & Software \\
\hline
Bolt       & 25-cm SCT & ST-7  & MuniPack$^a$ \\
Nelson     & 32-cm reflector & ST-8E & AIP4Win \\
Monard     & 30-cm SCT & ST-7E & AIP4Win \\
\hline
 \multicolumn{4}{l}{$^a$ http://munipack.astronomy.cz.} \\
\end{tabular}
\end{center}
\end{table}

\section{Observations}

\subsection{CCD Observations}

   The observers, equipment and reduction software are summarized in
Table \ref{tab:equipment}.  All observers performed aperture photometry
implemented in the packages listed in Table \ref{tab:equipment}.
The observations used unfiltered CCD systems having a response
close to Kron-Cousins $R_{\rm c}$ band for outbursting dwarf novae.
The errors of single measurements are typically
less than 0.01--0.03 mag.  The magnitudes were determined relative to
GSC 8347.944, whose constancy during the observation was confirmed by
a comparison with GSC 2.2 S230002154022.  The relative magnitudes by
PN using the primary comparison star of GSC 8347.1475 have been converted
to the common scale by adding a constant of $-$0.691 mag.

   Barycentric corrections to the observed times were applied before the
following analysis.

\subsection{Visual Observations}

   Visual observations were done with 32-cm (RS), 40-cm (AP),
32-cm (PN) and 32-cm (BM) reflectors.  All observations were done using
photoelectrically calibrated $V$-magnitude comparison stars.  The typical
error of visual estimates was 0.2 mag.  The observations were used to
determine the outburst cycle lengths and characteristics.
CCD monitoring observation by TR (18-cm refractor and an unfiltered ST-7E)
has been included in the analysis.

\begin{table}
\caption{Journal of the 2002 CCD photometry of BF Ara.}\label{tab:log}
\begin{center}
\begin{tabular}{crccrc}
\hline\hline
\multicolumn{2}{c}{2002 Date}& Start--End$^a$ & Exp(s) & $N$
        & Obs$^b$ \\
\hline
August & 18 & 52504.961--52505.055 &  90 &  76 & N \\
       & 18 & 52504.985--52505.210 & 30--45 & 384 & B \\
       & 19 & 52505.973--52506.228 &  60 & 317 & B \\
       & 20 & 52506.952--52507.230 &  60 & 339 & B \\
       & 21 & 52507.988--52508.224 &  60 & 297 & B \\
       & 22 & 52509.044--52509.100 & 240 &  22 & N \\
       & 23 & 52510.201--52510.442 &  50 & 312 & M \\
       & 25 & 52511.939--52512.092 & 210 &  67 & N \\
       & 26 & 52512.883--52513.069 & 180 &  69 & N \\
       & 27 & 52513.889--52514.026 & 200 &  53 & N \\
\hline
 \multicolumn{6}{l}{$^a$ BJD$-$2400000.} \\
 \multicolumn{6}{l}{$^b$ N (Nelson), B (Bolt), M (Monard)} \\
\end{tabular}
\end{center}
\end{table}

\begin{figure}
  \includegraphics[angle=0,width=8.8cm]{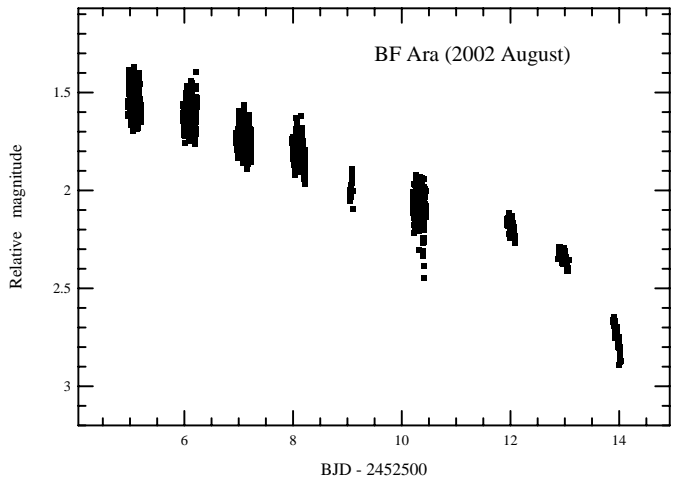}
  \caption{Light curve of the 2002 August superoutburst of BF Ara.
  The magnitudes are given relative to GSC 8347.944 (approximate USNO A2.0
  $r$ magnitude 12.4), and are on a system close to $R_{\rm c}$.}
  \label{fig:lc}
\end{figure}

\begin{figure}
  \includegraphics[angle=0,width=8.8cm,height=11cm]{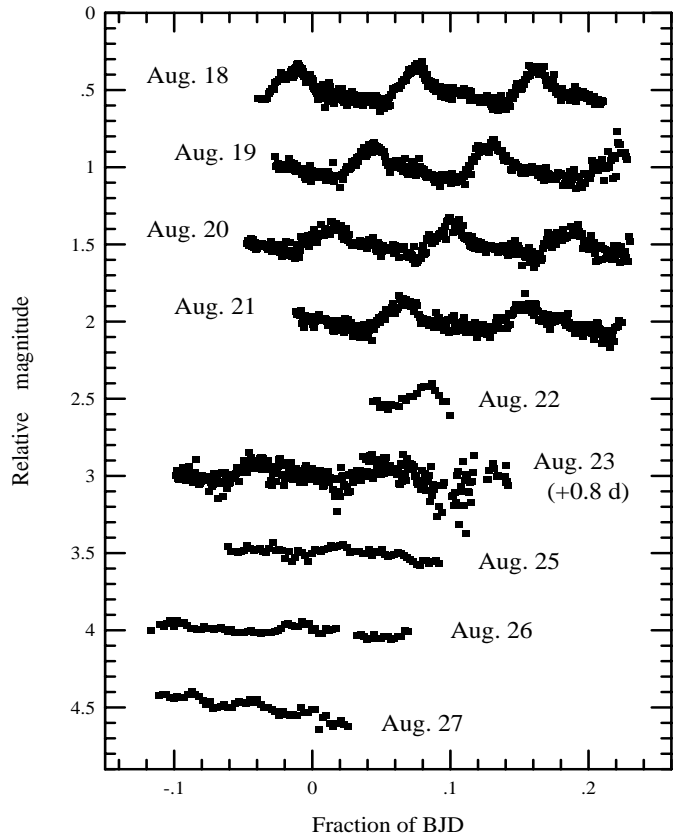}
  \caption{Nightly light curves of BF Ara.  Superhumps are clearly visible.}
  \label{fig:night}
\end{figure}

\section{The 2002 August Superoutburst}

\subsection{Course of Outburst}\label{sec:course}

   The 2002 August outburst was detected by RS on August 14.415 UT at
a visual magnitude of 14.4.  The object was reported to be fainter than
14.8 on August 13.491 UT.  The object further brightened to a magnitude
of 14.0 on August 16.478 UT.  Because the outburst was apparently
a long, bright outburst (likely superoutburst), we initiated a CCD
photometric campaign through the VSNET Collaboration.  From the August 18
observations by GB and PN, unmistakable superhumps were detected
(vsnet-alert 7450)\footnote{
http://www.kusastro.kyoto-u.ac.jp/vsnet/Mail/alert7000/\\msg00450.html.
}, qualifying BF Ara as a genuine SU UMa-type dwarf nova (see section
\ref{sec:SH} for more details).  The journal
of the CCD observations is listed in Table \ref{tab:log}.

   Figure \ref{fig:lc} shows the light curve of the outburst based on
CCD photometry.  The slowly fading (0.10 mag d$^{-1}$) superoutburst
plateau phase and a more rapidly fading phase on August 27 (BJD 2452514)
are clearly demonstrated.  The plateau phase lasted for 13 d
since the start of the outburst.  Nightly light curves are presented in
Figure \ref{fig:night}, demonstrating the clear presence of superhumps.

\begin{figure}
  \includegraphics[angle=0,width=8cm]{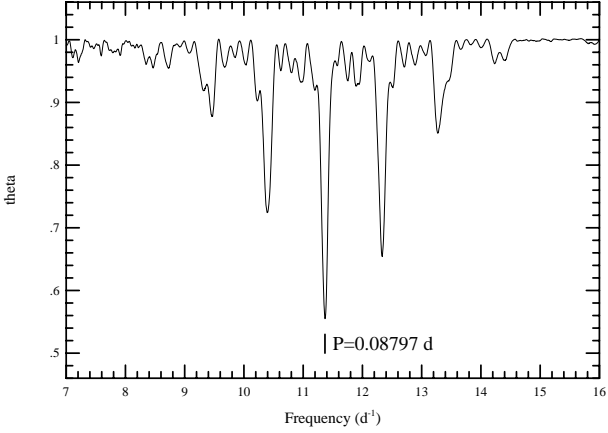}
  \caption{Period analysis of BF Ara.  The strongest signal at a frequency
  of 11.368(2) d$^{-1}$ corresponds to a mean superhump period of
  $P_{\rm SH}$ = 0.08797(1) d.}
  \label{fig:pdm}
\end{figure}

\begin{figure}
  \includegraphics[angle=0,width=8cm]{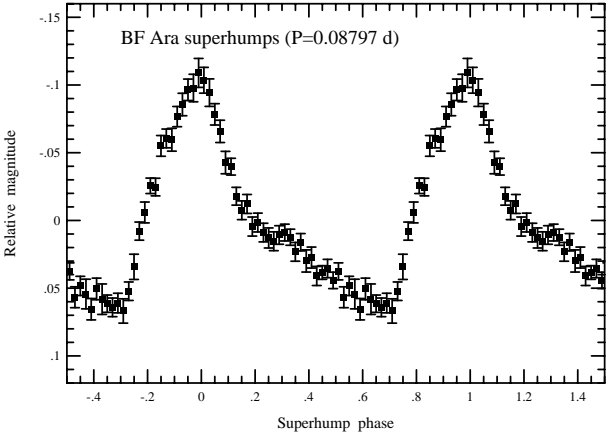}
  \caption{Mean superhump profile of BF Ara.}
  \label{fig:ph}
\end{figure}

\subsection{Superhump Period}\label{sec:SH}

   Figure \ref{fig:pdm} shows the result of a period analysis using
Phase Dispersion Minimization (PDM: \citealt{PDM}) applied to the data set
covering the superoutburst plateau (2002 August 18--26), after removing
the linear decline trend.  The best determined frequency of superhumps
is 11.368(2) d$^{-1}$, corresponding to a mean superhump period of
$P_{\rm SH}$ = 0.08797(1) d.  The significance of this period is
better than 99.99\%.

Figure \ref{fig:ph} shows the phase-averaged profile of superhumps
at the period of 0.08797 d.  The rapidly rising and slowly
fading superhump profile is very characteristic of an SU UMa-type
dwarf nova \citep{vog80suumastars,war85suuma}.

\begin{table}
\caption{Times of superhump maxima of BF Ara.}\label{tab:max}
\begin{center}
\begin{tabular}{ccc}
\hline\hline
$E^a$  & BJD$-$2400000 & $O-C^b$ \\
\hline
 0     & 52504.9903 &  0.0001 \\
 1     & 52505.0762 & $-$0.0019 \\
 2     & 52505.1627 & $-$0.0033 \\
12     & 52506.0453 &  0.0001 \\
13     & 52506.1314 & $-$0.0017 \\
14     & 52506.2231 &  0.0021 \\
23     & 52507.0149 &  0.0026 \\
24     & 52507.1034 &  0.0032 \\
25     & 52507.1898 &  0.0017 \\
35     & 52508.0682 &  0.0009 \\
36     & 52508.1557 &  0.0005 \\
46.5$^c$ & 52509.0837 &  0.0054 \\
60     & 52510.2636 & $-$0.0016 \\
61     & 52510.3512 & $-$0.0019 \\
80     & 52512.0202 & $-$0.0033 \\
90     & 52512.9027 &  0.0000 \\
91     & 52512.9924 &  0.0018 \\
101.5$^d$ & 52513.9127 & $-$0.0011 \\
102    & 52513.9587 &  0.0010 \\
102.5$^d$ & 52513.9995 & $-$0.0022 \\
\hline
 \multicolumn{3}{l}{$^a$ Cycle count since BJD 2452504.990.} \\
 \multicolumn{3}{l}{$^b$ $O-C$ calculated against equation
                    \ref{equ:reg1}.} \\
 \multicolumn{3}{l}{$^c$ Likely secondary superhump maximum.} \\
 \multicolumn{3}{l}{$^d$ Late superhumps.} \\
\end{tabular}
\end{center}
\end{table}

\begin{figure}
  \includegraphics[angle=0,width=8cm]{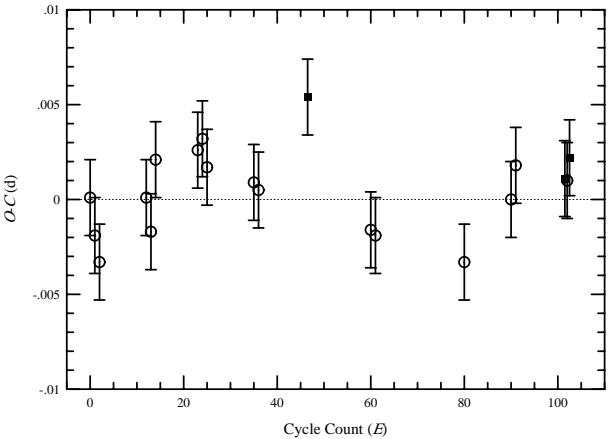}
  \caption{$O-C$ diagram of superhump maxima of BF Ara.  The error
  bars correspond to the upper limits of the errors.
  The open circles denote (usual) superhumps.  The filled squares represent
  secondary superhump maximum ($E$ = 46.5) and late superhumps
  ($E$ = 101.5 and $E$ = 102.5).  The $O-C$'s are virtually zero,
  indicating an exceptionally small period derivative.
  }
  \label{fig:oc}
\end{figure}

\subsection{Superhump Period Change}

   We extracted the maximum times of superhumps from the light curve by eye.
The averaged times of a few to several points close to the maxima were
used as representatives of the maximum times.  The errors of the maximum
times are less than $\sim$0.002 d.  The resultant superhump maxima
are given in Table \ref{tab:max}.  The values are given to 0.0001 d in
order to avoid the loss of significant digits in a later analysis.
The cycle count ($E$) is defined as the cycle number since BJD 2452504.990.
The maximum at $E$ = 46.5 likely corresponds to a secondary superhump
maximum, which is sometimes observed around superhump phases at 0.4--0.6
\citep{uda90suuma,kat92swumasuperQPO}.  The maxima at $E$ = 101.5 and
102.5 correspond to late superhumps
\citep{hae79lateSH,vog83lateSH,vanderwoe88lateSH,hes92lateSH},
which are known to have similar periods with ordinary superhumps, but
have phases of $\sim$0.5 different from those of ordinary superhumps.
Excluding the maxima of the likely secondary superhump and late superhumps,
a linear regression to the observed superhump times gives the following
ephemeris (the errors correspond to 1$\sigma$ errors at $E$ = 39):

\begin{equation}
{\rm BJD (maximum)} = 2452504.9902(5) + 0.087917(15) E. \label{equ:reg1}
\end{equation}

   The derived $O-C$'s against equation \ref{equ:reg1} are almost zero
within the expected errors of the maximum times
(Figure \ref{fig:oc}).
A quadratic fit yielded a period derivative of
$\dot{P}$ = $-$0.6$\pm$1.2 $\times$ 10$^{-6}$ d cycle$^{-1}$,
or $P_{\rm dot}$ = $\dot{P}/P$ = $-$0.8$\pm$1.4 $\times$ 10$^{-5}$.
This virtually zero period derivative makes a clear contrast against
recently discovered SU UMa-type dwarf novae (V877 Ara, KK Tel), which
have large negative period derivatives \citep{kat03v877arakktelpucma}.

\section{Astrometry and Quiescent Identification}

   The quiescent counterpart of BF Ara has been suggested by
\citet{vog82atlas}.  Since this field is very crowded, we have tried
to make an unambiguous independent identification based on outburst
CCD images.

\begin{figure}
  \begin{center}
  \includegraphics[angle=0,width=8cm]{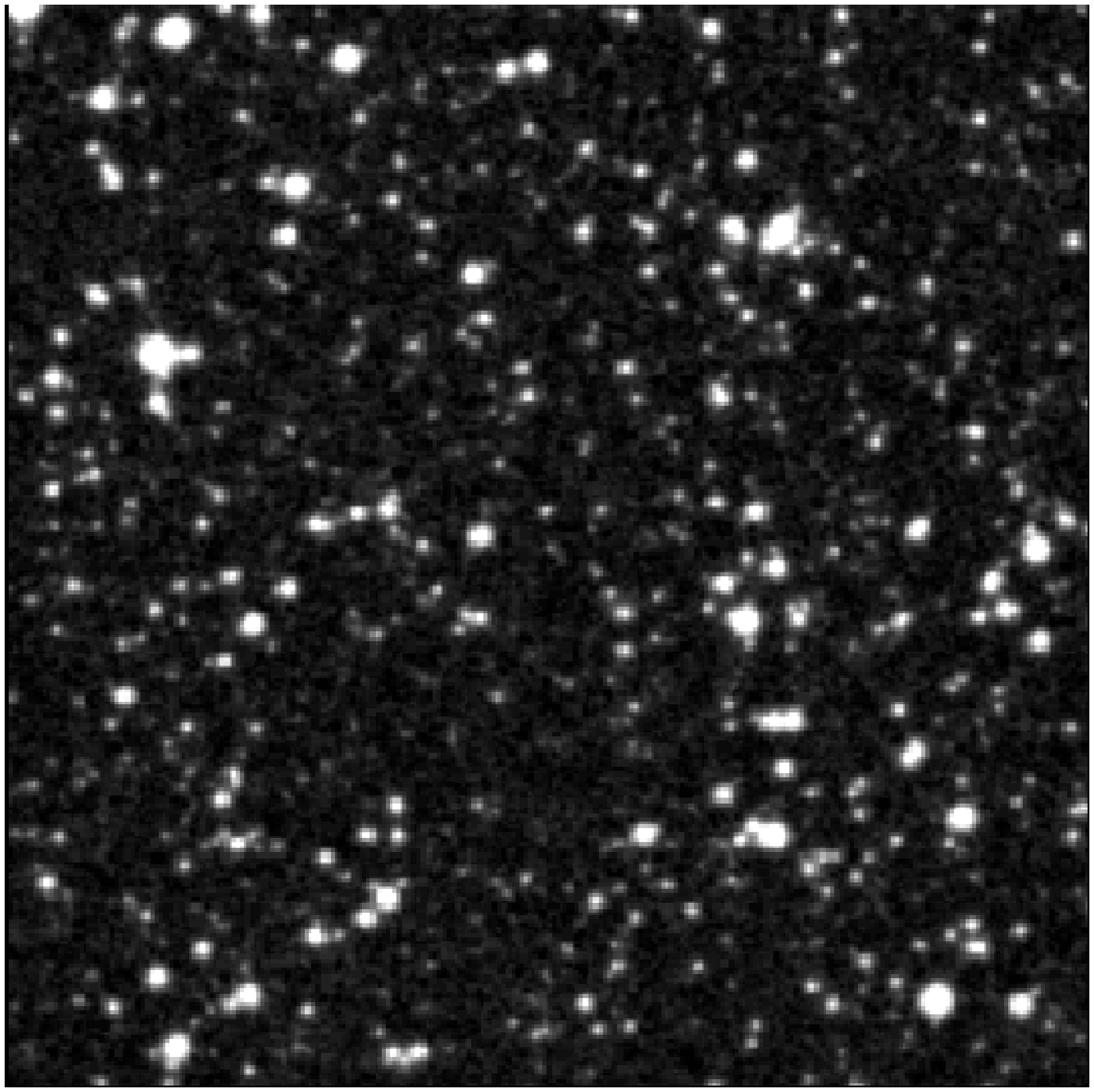} \\
  \vskip 1mm
  \includegraphics[angle=0,width=8cm]{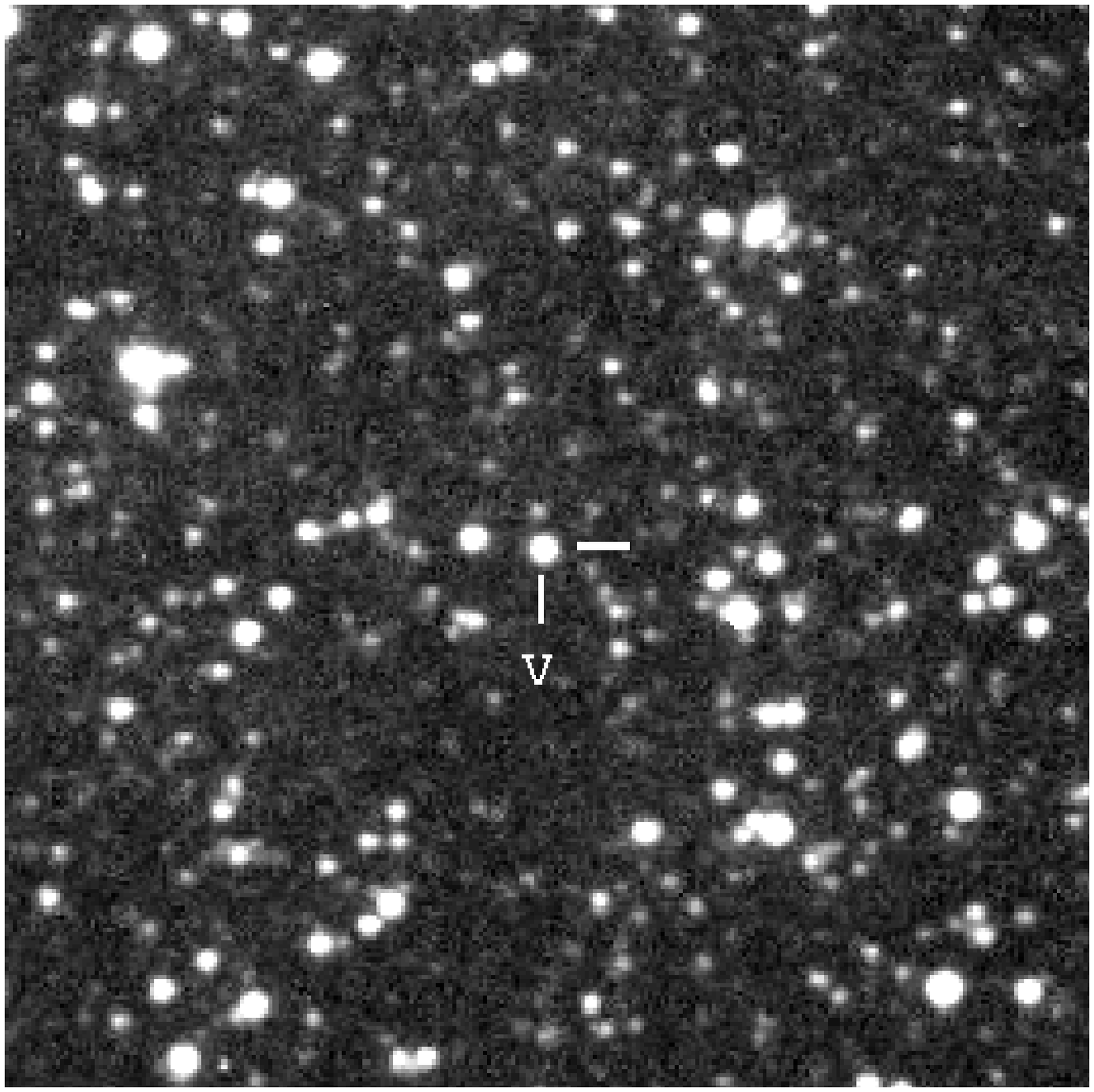}
  \end{center}
  \caption{Identification of BF Ara.  5 arcminutes square, north is up
  and left is east for each image.  (Upper:) In quiescence,
  reproduced from the DSS 2 red image taken on 1996 Sept. 6.  
  (Lower:) In outburst, taken on 2002 August 18 by PN.  V = BF Ara.}
  \label{fig:id}
\end{figure}

   Astrometry of the outbursting BF Ara was performed on CCD
images taken by GB and PN.  An average of measurements
of seven images (UCAC1 system, 60 -- 240 reference stars; internal
dispersion of the measurements was $\sim$0$''$.08) has yielded
a position of 17$^h$ 38$^m$ 21$^s$.322, $-$47$^{\circ}$ 10$'$ 41$''$.46
(J2000.0).  The position agrees with the GSC$-$2.2.1 star at
17$^h$ 38$^m$ 21$^s$.307, $-$47$^{\circ}$ 10$'$ 41$''$.10
(epoch 1996.680 and magnitude $r$ = 17.67), which is most
likely the quiescent counterpart of BF Ara (Figure \ref{fig:id}).
Comparing with our position and the oldest DSS image (epoch =
1979.367), no apparent proper motion was detected; its upper 
limit is deduced to be 0$''$.03 yr$^{-1}$.  Note that the DSS red
image taken on 1998 June 17 happened to catch BF Ara in outburst.

   The Astrographic Catalog (AC) contains an object about 2 arcseconds
from BF Ara.  The position in the latest version of AC is 
17$^h$ 38$^m$ 21$^s$.376, $-$47$^{\circ}$ 10$'$ 39$''$.31 (J2000.0,
epoch=1904.644 and magnitude $b$=13.11).  In case it was really BF Ara
in outburst, the deduced proper motion is $\sim$0$''$.022 yr$^{-1}$.

\begin{figure*}
  \includegraphics[angle=0,width=15cm]{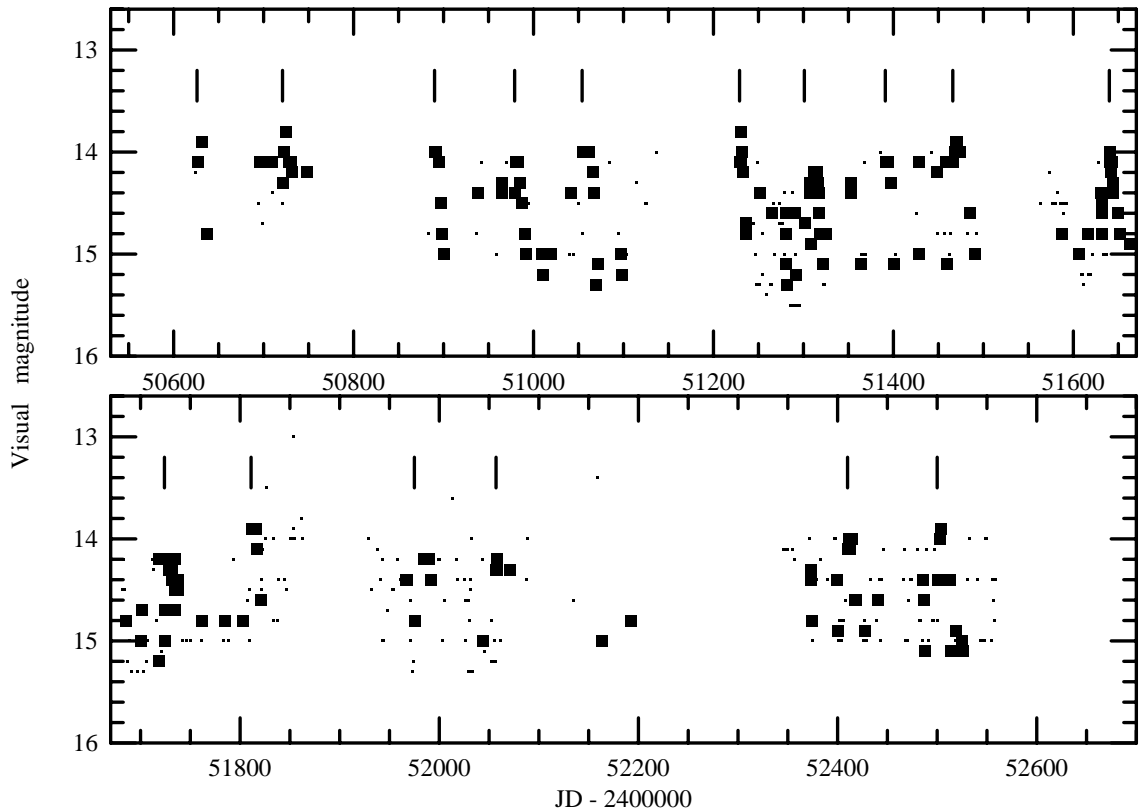}
  \caption{Long-term visual light curve of BF Ara.  The large and small dots
  represent positive and negative (upper limit) observations, respectively.
  The superoutbursts (see Table \ref{tab:out}) are marked with the vertical
  ticks.}
  \label{fig:long}
\end{figure*}

\begin{table}
\caption{List of Outbursts of BF Ara.}\label{tab:out}
\begin{center}
\begin{tabular}{ccccc}
\hline\hline
JD start$^a$ & JD end$^a$ & Max & Duration (d) & Type \\
\hline
50626.9 & 50636.9 & 13.9 & $>$10 & super \\
50695.9 & 50696.9 & 14.1 & 2 & normal \\
50707.9 & 50708.9 & 14.1 & 2 & normal \\
50721.9 & 50731.9 & 13.8 & 10 & super \\
50748.0 &   --    & 14.2 & 1 & normal \\
50890.3 & 50900.2 & 14.0 & $>$11 & super \\
50937.9 &   --    & 14.4 & 1: & normal \\
50965.0 &   --    & 14.3 & 1: & normal \\
50979.9 & 50992.1 & 14.1 & 13 & super \\
51008.9 & 51010.0 & 15.0 & 2 & normal \\
51041.9 &   --    & 14.4 & 1: & normal \\
51054.9 & 51071.9 & 14.0 & 17 & super \\
51096.9 & 51097.9 & 15.0 & 2 & normal \\
51229.2 & 51236.3 & 13.8 & $>$7 & super \\
51251.2 &   --    & 14.4 & 1 & normal \\
51265.3 &   --    & 14.6 & 1 & normal \\
51280.1 & 51281.3 & 14.6 & 2 & normal \\
51290.1 & 51291.2 & 14.6 & 2 & normal \\
51301.2 & 51321.3 & 14.2 & 17 & super \\
51325.3 &   --    & 14.8 & 1: & normal \\
51353.0 & 51353.3 & 14.3 & 1: & normal \\
51363.9 &   --    & 15.1 & 1: & normal \\
51391.9 & 51400.9 & 14.1 & $>$9 & super \\
51428.0 & 51428.9 & 14.1 & 1: & normal \\
51447.9 &   --    & 14.2 & 1: & normal \\
51458.0 & 51460.0 & 14.1 & 2 & normal \\
51466.0 & 51473.9 & 13.9 & $>$8 & super \\
51484.9 &   --    & 14.6 & 1: & normal$^b$ \\
51490.9 &   --    & 15.0 & 1: & normal \\
51587.1 &   --    & 14.8 & 1 & normal \\
51606.3 &   --    & 15.0 & 1: & normal \\
51616.3 &   --    & 14.8 & 1: & normal \\
51631.1 & 51632.3 & 14.4 & 2 & normal \\
51640.3 & 51651.3 & 14.0 & $>$11 & super \\
51663.1 &   --    & 14.9 & 1: & normal \\
51685.2 &   --    & 14.8 & 1 & normal \\
51700.0 & 51701.1 & 14.7 & 2 & normal \\
51718.3 & 51719.0 & 14.2 & 1 & normal \\
51724.9 & 51738.0 & 14.2 & $>$14 & super \\
51762.0 &   --    & 14.8 & 1: & normal \\
51785.0 &   --    & 14.8 & 1: & normal \\
51802.9 &   --    & 14.8 & 1: & normal \\
51811.9 & 51821.0 & 13.9 & $>$9 & super \\
51966.1 & 51967.3 & 14.4 & 2 & normal \\
51975.2 & 51992.1 & 14.2 & 17 & super \\
52044.0 &   --    & 15.0 & 1 & normal \\
52057.1 & 52070.9 & 14.2 & 13 & super \\
52163.0 &   --    & 15.0: & 1: & normal \\
52192.0 &   --    & 14.8 & 1: & normal \\
52373.1 & 52374.1 & 14.3 & 1 & normal \\
52399.0 & 52400.0 & 14.4 & 1 & normal \\
52410.2 & 52418.3 & 14.0 & $>$8 & super \\
52428.0 &   --    & 14.9 & 1 & normal \\
52440.2 &   --    & 14.6 & 1: & normal \\
52485.9 & 52487.9 & 14.4 & 2 & normal \\
52500.9 & 52513.9 & 13.9 & 13 & super \\
52519.0 &   --    & 14.9 & 1 & normal \\
52524.9 & 52525.9 & 15.0 & 2 & normal \\
\hline
 \multicolumn{5}{l}{$^a$ JD$-$2400000.} \\
 \multicolumn{5}{l}{$^b$ Continuation of the preceding superoutburst?} \\
\end{tabular}
\end{center}
\end{table}

\section{BF Ara as an SU UMa-type Dwarf Nova}

   Our observations have clearly established that BF Ara is indeed
an SU UMa-type dwarf nova.  This classification finally enables us to
unambiguously determine the outburst types and supercycle length.
Figure \ref{fig:long} shows a long-term light curve covering the interval
1997 June -- 2002 October.

   Table \ref{tab:out} lists the recorded outbursts of BF Ara.
The table is an updated extension of the table in \citet{kat01bfara},
who only listed superoutburst candidates which had been recorded at that
time.  The durations generally correspond to the durations when the variable
was brighter than 15.0 mag.  The durations of single or a few solitary
observations have been supplemented with an uncertainty mark (:).
The durations of outbursts have a clear bimodal ($\leq$2 d or $\geq$7 d)
distribution, which is very characteristic of an SU UMa-type star
\citep{vog80suumastars,war85suuma}.  The type identification of the
outbursts was primarily based on their durations.

\begin{figure}
  \includegraphics[angle=0,width=8cm]{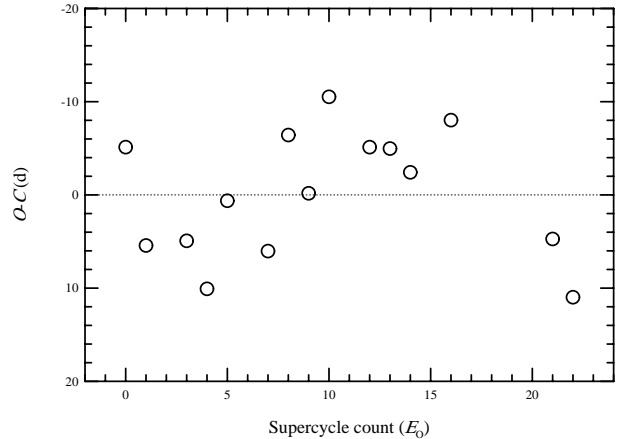}
  \caption{$O-C$ diagram of supermaxima of BF Ara.
  The $O-C$'s were calculated against equation \ref{equ:soreg1}.
  }
  \label{fig:sooc}
\end{figure}

   Since the occurrence of superoutbursts is quite regular
\citep{kat01bfara}, we have made a reanalysis of the supercycle length based
on the new material.
A linear regression to the observed start times of supermaxima gives the
following ephemeris:

\begin{equation}
{\rm JD (maximum)} = 2450632.4 + 84.34 E_O, \label{equ:soreg1}
\end{equation}

  where $E_O$ denotes the number of supercycles since the first (JD
2450626.9) superoutburst.  The refined mean supercycle length is 84.3(3) d.
The $O-C$'s against this equation are displayed in Figure \ref{fig:sooc}.
The $|O-C|$'s are usually within 10 d.  The small $|O-C|$ values are
almost comparable to those of ER UMa stars
\citep{hon95erumarzlmiv1159oriv446her,rob95eruma,kat01v1159ori},
although the short-term stability of the supercycle is not as marked as
in ER UMa stars.

\section{Relation to ER UMa Stars}

   In addition to short $T_s$, ER UMa stars have distinct outburst
properties.  They can be summarized as: (1) extremely short ($\sim$4 d)
recurrence time of normal outbursts,
(2) extremely large (0.30--0.45) duty cycles of superoutbursts
(see folded figures in \citealt{rob95eruma,kat01v1159ori})
(3) low outburst amplitudes (2--3 mag).  These properties are the natural
consequences from the disk-instability model in high-$\dot{M}$ systems
\citep{osa95eruma}.  These properties can thus be reasonably used
to discriminate ER UMa stars from (a larger population of) SU UMa-type
dwarf novae.

   In the case of BF Ara, the shortest observed intervals
(see Table \ref{tab:out}) of normal outbursts was 6 d, although most of
the shortest intervals are close to 10 d.  These values more resemble
those of usual SU UMa-type
dwarf novae with the shortest recurrence times.\footnote{
  The exact conclusion would await further dense, deep observations,
  since some of the faint, short, normal outbursts could have easily
  escaped from detection in the present study.
}  The durations of the well-observed superoutbursts were typically
11--17 d (Table \ref{tab:out}).  The detailed CCD observation of the
2002 August superoutburst (the duration being 13 d)
is in agreement with these estimates.  These values correspond to
superoutburst duty cycles of 0.13--0.20, which are noticeably smaller
than those of ER UMa stars.

   Some properties of the superoutburst of BF Ara are also unlike
those of ER UMa stars.
The mean decline rate (0.10 mag d$^{-1}$) of the superoutburst plateau
(cf. section \ref{sec:course}) is also close to those of usual SU UMa-type
stars \citep{kat02v359cen}, rather than an extremely small value of
$\sim$0.04 mag d$^{-1}$ in ER UMa \citep{kat95eruma}.  The evolution
of superhumps (section \ref{sec:SH}) is also quite normal for a usual
SU UMa-type dwarf nova with smoothly decaying amplitudes of superhumps,
in contrast to ER UMa stars which show a rapid initial decay of
the superhump amplitudes and a later regrowth \citep{kat96erumaSH}.

   These features indicate that BF Ara should be classified as a usual
SU UMa-type dwarf nova rather than an ER UMa star.  BF Ara is thus
qualified as a usual SU UMa-type dwarf nova with the shortest
measured $T_s$.  Despite the past and present intensive studies of the
most promising candidates of transitional objects, there still remains
an unfilled gap of distributions between ER UMa stars and usual
SU UMa-type dwarf novae.

\section{On the Superhump Period Change}

   The periods of ``textbook'' superhumps in usual SU UMa-type dwarf
novae are known to decrease at a rather common rate of
$\dot{P}/P \sim -5 \times 10^{-5}$ during superoutbursts
(e.g. \citealt{war85suuma,pat93vyaqr}; for a recent progress, see
\citealt{kat03v877arakktelpucma}).  This decrease of the superhump periods
has usually been attributed to a decrease in the angular velocity of
precession of an eccentric disk, which is caused by a decrease in the
disk radius during superoutbursts \citep{osa85SHexcess}.

   In recent years, however, several systems have
been found to show zero to positive (increase of the periods) period
derivatives.  The best-established examples include WZ Sge-type dwarf
novae (SU UMa-type dwarf novae with very infrequent (super)outbursts,
see e.g. \citealt{bai79wzsge,dow81wzsge,pat81wzsge,odo91wzsge,kat01hvvir})
and related large-amplitude systems (V1028 Cyg: \citealt{bab00v1028cyg};
SW UMa: \citealt{sem97swuma,nog98swuma}; WX Cet: \citealt{kat01wxcet}).
Since all of these objects have short orbital periods, small $q$,
and small $\dot{M}$, there has been a suggestion that either $q$ or
low $\dot{M}$ is responsible for the period increase \citep{kat01hvvir}.
The most recent discoveries of long-period (thus likely large $q$),
and likely low-$\dot{M}$ SU UMa-type dwarf novae (V725 Aql:
\citealt{uem01v725aql}; EF Peg: K. Matsumoto et al, in preparation, see
also \citealt{kat02efpeg}), having zero or marginally positive $P_{\rm dot}$,
have more preferred the interpretation requiring a low $\dot{M}$.

   The present discovery of a virtually zero $\dot{P}$ in a long-period
($P_{\rm SH}$ = 0.08797(1) d), otherwise relatively normal, system is
therefore surprising.  This discovery has not only strengthened the
previously neglected diversity of $P_{\rm dot}$ in long-period SU UMa-type
systems claimed in \citet{kat03v877arakktelpucma}, but also provides
a new clue to understand the physics of superhump period changes.

   Since BF Ara has short outburst recurrence times (both superoutbursts
and normal outbursts), $\dot{M}$ is expected to high \citep{ich94cycle}.
By using typical supercycles of BF Ara (84.3 d) and V725 Aql ($\sim$900 d),
the expected $\dot{M}$ in BF Ara is $\sim$10 times larger than that
in V725 Aql \citep{ich94cycle}.
The occurrence of nearly zero $P_{\rm dot}$ systems
in a wide region of $\dot{M}$ implies that the interpretation requiring
a low $\dot{M}$ would be no longer valid, or a different mechanism is
responsible for BF Ara.  \citet{mur00SHprecession,mon01SH}
recently suggested that the (prograde) apsidal motion of the eccentric
disk can be reduced by introducing pressure forces.
A high $\dot{M}$ in BF Ara may have
modified the usual time-evolution of superhump period through this
pressure effect.  If this is the case, we can expect a more prominent
effect in ER UMa stars, although a limited $P_{\rm dot}$ measurement
\citep{pat95v1159ori} failed to allow us a definitive conclusion.
Further observations of $P_{\rm dot}$ in more systems with a wide range
of parameters are definitely needed.

\section*{Acknowledgments}

This work is partly supported by a grant-in-aid [13640239 (TK),
14740131 (HY)] from the Japanese Ministry of Education, Culture, Sports,
Science and Technology.
The CCD operation of the Bronberg Observatory is partly sponsored by
the Center for Backyard Astrophysics.
The CCD operation by Peter Nelson is on loan from the AAVSO,
funded by the Curry Foundation.
This research has made use of the Digitized Sky Survey producted by STScI, 
the ESO Skycat tool, the VizieR catalogue access tool.

\label{lastpage}


\begin{thebibliography}{}

\bibitem[\protect\citeauthoryear{Baba, Kato, Nogami, Hirata, Matsumoto \&
  Sadakane}{Baba et~al.}{2000}]{bab00v1028cyg}
Baba H.,  Kato T.,  Nogami D.,  Hirata R.,  Matsumoto K.,    Sadakane K.,
  2000, PASJ, 52, 429

\bibitem[\protect\citeauthoryear{Bailey}{Bailey}{1979}]{bai79wzsge}
Bailey J.,  1979, MNRAS, 189, 41P

\bibitem[\protect\citeauthoryear{Bruch}{Bruch}{1983}]{bru83bfara}
Bruch A.,  1983, Inf. Bull. Var. Stars, 2286

\bibitem[\protect\citeauthoryear{Buat-M\'{e}nard \& Hameury}{Buat-M\'{e}nard \&
  Hameury}{2002}]{bua02suumamodel}
Buat-M\'{e}nard V.,  Hameury J.-M.,  2002, A\&A, 386, 891

\bibitem[\protect\citeauthoryear{Buat-M\'{e}nard, Hameury \&
  Lasota}{Buat-M\'{e}nard et~al.}{2001}]{bua01DNoutburst}
Buat-M\'{e}nard V.,  Hameury J.-M.,    Lasota J.-P.,  2001, A\&A, 366, 612

\bibitem[\protect\citeauthoryear{Downes \& Margon}{Downes \&
  Margon}{1981}]{dow81wzsge}
Downes R.~A.,  Margon B.,  1981, MNRAS, 197, 35P

\bibitem[\protect\citeauthoryear{Haefner, Schoembs \& Vogt}{Haefner
  et~al.}{1979}]{hae79lateSH}
Haefner R.,  Schoembs R.,    Vogt N.,  1979, A\&A, 77, 7

\bibitem[\protect\citeauthoryear{Harvey, Skillman, Patterson \&
  Ringwald}{Harvey et~al.}{1995}]{har95v503cyg}
Harvey D.,  Skillman D.~R.,  Patterson J.,    Ringwald F.~A.,  1995, PASP,
  107, 551

\bibitem[\protect\citeauthoryear{Hellier}{Hellier}{2001}]{hel01eruma}
Hellier C.,  2001, PASP, 113, 469

\bibitem[\protect\citeauthoryear{Hessman, Mantel, Barwig \& Schoembs}{Hessman
  et~al.}{1992}]{hes92lateSH}
Hessman F.~V.,  Mantel K.-H.,  Barwig H.,    Schoembs R.,  1992, A\&A, 263, 147

\bibitem[\protect\citeauthoryear{Honeycutt, Robertson \& Turner}{Honeycutt
  et~al.}{1995}]{hon95erumarzlmiv1159oriv446her}
Honeycutt R.~K.,  Robertson J.~W.,    Turner G.~W.,  1995, in Bianchini A.,
  della Valle M.,   Orio M.,  eds, Cataclysmic Variables
  (Dordrecht: Kluwer Academic Publishers), p.~75

\bibitem[\protect\citeauthoryear{Ichikawa \& Osaki}{Ichikawa \&
  Osaki}{1994}]{ich94cycle}
Ichikawa S.,  Osaki Y.,  1994, in Duschl W.~J.,  Frank J.,  Meyer F.,
  Meyer-Hofmeister E.,   Tscharnuter W.~M.,  eds, Theory of Accretion Disks-2
  (Dordrecht: Kluwer Academic Publishers), p.~169

\bibitem[\protect\citeauthoryear{Ishioka, Kato, Uemura, Iwamatsu, Matsumoto,
  Martin, Billings \& Novak}{Ishioka et~al.}{2001}]{ish01ixdra}
Ishioka R.,  Kato T.,  Uemura M.,  Iwamatsu H.,  Matsumoto K.,  Martin B.~E.,
  Billings G.~W.,    Novak R.,  2001, PASJ, 53, L51

\bibitem[\protect\citeauthoryear{Kato}{Kato}{2001}]{kat01v1159ori}
Kato T.,  2001, PASJ, 53, L17

\bibitem[\protect\citeauthoryear{Kato}{Kato}{2002}]{kat02efpeg}
Kato T.,  2002, PASJ, 54, 87

\bibitem[\protect\citeauthoryear{Kato, Hanson, Poyner, Muyllaert, Reszelski \&
  Dubovsky}{Kato et~al.}{2000}]{kat00ssumi}
Kato T.,  Hanson G.,  Poyner G.,  Muyllaert E.,  Reszelski M.,    Dubovsky
  P.~A.,  2000, Inf. Bull. Var. Stars, 4932

\bibitem[\protect\citeauthoryear{Kato, Hirata \& Mineshige}{Kato
  et~al.}{1992}]{kat92swumasuperQPO}
Kato T.,  Hirata R.,    Mineshige S.,  1992, PASJ, 44, L215

\bibitem[\protect\citeauthoryear{Kato, Ishioka \& Uemura}{Kato
  et~al.}{2002}]{kat02v503cyg}
Kato T.,  Ishioka R.,    Uemura M.,  2002, PASJ, 54, 1029

\bibitem[\protect\citeauthoryear{Kato \& Kunjaya}{Kato \&
  Kunjaya}{1995}]{kat95eruma}
Kato T.,  Kunjaya C.,  1995, PASJ, 47, 163

\bibitem[\protect\citeauthoryear{Kato, Lipkin, Retter \& Leibowitz}{Kato
  et~al.}{1998}]{kat98ssumi}
Kato T.,  Lipkin Y.,  Retter A.,    Leibowitz E.,  1998,
  Inf. Bull. Var. Stars, 4602

\bibitem[\protect\citeauthoryear{Kato, Matsumoto, Nogami, Morikawa \&
  Kiyota}{Kato et~al.}{2001}]{kat01wxcet}
Kato T.,  Matsumoto K.,  Nogami D.,  Morikawa K.,    Kiyota S.,  2001, PASJ,
  53, 893

\bibitem[\protect\citeauthoryear{Kato, Nogami \& Baba}{Kato
  et~al.}{1996}]{kat96diuma}
Kato T.,  Nogami D.,    Baba H.,  1996, PASJ, 48, L93

\bibitem[\protect\citeauthoryear{Kato, Nogami, Baba, Masuda, Matsumoto \&
  Kunjaya}{Kato et~al.}{1999}]{kat99erumareview}
Kato T.,  Nogami D.,  Baba H.,  Masuda S.,  Matsumoto K.,    Kunjaya C.,  1999,
  in Mineshige S.,  Wheeler J.~C.,  eds, Disk Instabilities in Close Binary
  Systems (Tokyo: Universal Academy Press), p.~45

\bibitem[\protect\citeauthoryear{Kato, Nogami \& Masuda}{Kato
  et~al.}{1996}]{kat96erumaSH}
Kato T.,  Nogami D.,    Masuda S.,  1996, PASJ, 48, L5

\bibitem[\protect\citeauthoryear{Kato, Nogami, Masuda \& Baba}{Kato
  et~al.}{1998}]{kat98hsvir}
Kato T.,  Nogami D.,  Masuda S.,    Baba H.,  1998, PASP, 110, 1400

\bibitem[\protect\citeauthoryear{Kato, Nogami, Masuda \& Hirata}{Kato
  et~al.}{1995}]{kat95hsvir}
Kato T.,  Nogami D.,  Masuda S.,    Hirata R.,  1995, Inf. Bull. Var. Stars,
  4193

\bibitem[\protect\citeauthoryear{Kato, Santallo, Bolt, Richards, Nelson,
  Monard, Uemura, Kiyota, Stubbings, Pearce, Watanabe, Schmeer \& Yamaoka}{Kato
  et~al.}{2002}]{kat03v877arakktelpucma}
Kato T.,  Santallo S.,  Bolt G.,  Richards T.,  Nelson P.,  Monard B.,  Uemura
  M.,  Kiyota S.,  Stubbings R.,  Pearce A.,  Watanabe T.,  Schmeer P.,
  Yamaoka H.,  2003, MNRAS, in press (astro-ph/0210674)

\bibitem[\protect\citeauthoryear{Kato, Sekine \& Hirata}{Kato
  et~al.}{2001}]{kat01hvvir}
Kato T.,  Sekine Y.,    Hirata R.,  2001, PASJ, 53, 1191

\bibitem[\protect\citeauthoryear{Kato, Stubbings, Nelson, Santallo, Ishioka,
  Uemura, Sumi, Muraki, Kilmartin, Bond, Noda, Yock, Hearnshaw, Monard \&
  Yamaoka}{Kato et~al.}{2002}]{kat02v359cen}
Kato T.,  Stubbings R.,  Nelson P.,  Santallo R.,  Ishioka R.,  Uemura M.,
  Sumi T.,  Muraki Y.,  Kilmartin P.,  Bond I.,  Noda S.,  Yock P.,  Hearnshaw
  J.~B.,  Monard B.,    Yamaoka H.,  2002, A\&A, 395, 541

\bibitem[\protect\citeauthoryear{Kato, Stubbings, Pearce, Dubovsky, Kiyota,
  Itoh \& Simonsen}{Kato et~al.}{2001}]{kat01hsvir}
Kato T.,  Stubbings R.,  Pearce A.,  Dubovsky P.~A.,  Kiyota S.,  Itoh H.,
  Simonsen M.,  2001, Inf. Bull. Var. Stars, 5109

\bibitem[\protect\citeauthoryear{Kato, Stubbings, Pearce, Nelson \&
  Monard}{Kato et~al.}{2001}]{kat01bfara}
Kato T.,  Stubbings R.,  Pearce A.,  Nelson P.,    Monard B.,  2001,
  Inf. Bull. Var. Stars, 5119

\bibitem[\protect\citeauthoryear{Kholopov, Samus', Frolov, Goranskij, Gorynya,
  Kireeva, Kukarkina, Kurochkin, Medvedeva, Perova \& Shugarov}{Kholopov
  et~al.}{1985}]{GCVS}
Kholopov P.~N.,  Samus' N.~N.,  Frolov M.~S.,  Goranskij V.~P.,  Gorynya N.~A.,
   Kireeva N.~N.,  Kukarkina N.~P.,  Kurochkin N.~E.,  Medvedeva G.~I.,  Perova
  N.~B.,    Shugarov S.~Y.,  1985, General Catalogue of Variable Stars, fourth
  edition.
Moscow: Nauka Publishing House

\bibitem[\protect\citeauthoryear{King}{King}{1988}]{kin88binaryevolution}
King A.~R.,  1988, QJRAS, 29, 1

\bibitem[\protect\citeauthoryear{King}{King}{2000}]{kin00CVevolution}
King A.~R.,  2000, New Astronomy Reviews, 44, 167

\bibitem[\protect\citeauthoryear{King, Frank, Kolb \& Ritter}{King
  et~al.}{1995}]{kin95masstransfercycle}
King A.~R.,  Frank J.,  Kolb U.,    Ritter H.,  1995, ApJ, 444, L37

\bibitem[\protect\citeauthoryear{King, Frank, Kolb \& Ritter}{King
  et~al.}{1996}]{kin96masstransfercycle}
King A.~R.,  Frank J.,  Kolb U.,    Ritter H.,  1996, ApJ, 467, 761

\bibitem[\protect\citeauthoryear{Kolb, Rappaport, Schenker \& Howell}{Kolb
  et~al.}{2001}]{kol01mdotvar}
Kolb U.,  Rappaport S.,  Schenker K.,    Howell S.,  2001, ApJ, 563, 958

\bibitem[\protect\citeauthoryear{McCormick}{McCormick}{1998}]{mcc98masstransfe%
rcycle}
McCormick P.,  1998, in Howell S.,  Kuulkers E.,   Woodward C.,  eds,
  ASP Conf. Ser. 137, Wild Stars in the Old West
  (San Francisco: ASP), p.~415

\bibitem[\protect\citeauthoryear{Misselt \& Shafter}{Misselt \&
  Shafter}{1995}]{mis95PGCV}
Misselt K.~A.,  Shafter A.~W.,  1995, AJ, 109, 1757

\bibitem[\protect\citeauthoryear{Montgomery}{Montgomery}{2001}]{mon01SH}
Montgomery M.~M.,  2001, MNRAS, 325, 761

\bibitem[\protect\citeauthoryear{Murray}{Murray}{2000}]{mur00SHprecession}
Murray J.~R.,  2000, MNRAS, 314, 1P

\bibitem[\protect\citeauthoryear{Nogami, Baba, Kato \& Nov\'{a}k}{Nogami
  et~al.}{1998}]{nog98swuma}
Nogami D.,  Baba H.,  Kato T.,    Nov\'{a}k R.,  1998, PASJ, 50, 297

\bibitem[\protect\citeauthoryear{Nogami \& Kato}{Nogami \&
  Kato}{1997}]{nog97ciuma}
Nogami D.,  Kato T.,  1997, PASJ, 49, 109

\bibitem[\protect\citeauthoryear{Nogami, Kato, Masuda \& Hirata}{Nogami
  et~al.}{1995}]{nog95v1159ori}
Nogami D.,  Kato T.,  Masuda S.,    Hirata R.,  1995, Inf. Bull. Var. Stars,
  4155

\bibitem[\protect\citeauthoryear{Nogami, Kato, Masuda, Hirata, Matsumoto,
  Tanabe \& Yokoo}{Nogami et~al.}{1995}]{nog95rzlmi}
Nogami D.,  Kato T.,  Masuda S.,  Hirata R.,  Matsumoto K.,  Tanabe K.,
  Yokoo T.,  1995, PASJ, 47, 897

\bibitem[\protect\citeauthoryear{Nogami, Masuda \& Kato}{Nogami
  et~al.}{1997}]{nog97sxlmi}
Nogami D.,  Masuda S.,    Kato T.,  1997, PASP, 109, 1114

\bibitem[\protect\citeauthoryear{O'Donoghue, Chen, Marang, Mittaz, Winkler \&
  Warner}{O'Donoghue et~al.}{1991}]{odo91wzsge}
O'Donoghue D.,  Chen A.,  Marang F.,  Mittaz J. P.~D.,  Winkler H.,    Warner
  B.,  1991, MNRAS, 250, 363

\bibitem[\protect\citeauthoryear{Osaki}{Osaki}{1985}]{osa85SHexcess}
Osaki Y.,  1985, A\&A, 144, 369

\bibitem[\protect\citeauthoryear{Osaki}{Osaki}{1989}]{osa89suuma}
Osaki Y.,  1989, PASJ, 41, 1005

\bibitem[\protect\citeauthoryear{Osaki}{Osaki}{1995a}]{osa95eruma}
Osaki Y.,  1995a, PASJ, 47, L11

\bibitem[\protect\citeauthoryear{Osaki}{Osaki}{1995b}]{osa95rzlmi}
Osaki Y.,  1995b, PASJ, 47, L25

\bibitem[\protect\citeauthoryear{Osaki}{Osaki}{1996}]{osa96review}
Osaki Y.,  1996, PASP, 108, 39

\bibitem[\protect\citeauthoryear{Patterson}{Patterson}{1998}]{pat98evolution}
Patterson J.,  1998, PASP, 110, 1132

\bibitem[\protect\citeauthoryear{Patterson, Bond, Grauer, Shafter \&
  Mattei}{Patterson et~al.}{1993}]{pat93vyaqr}
Patterson J.,  Bond H.~E.,  Grauer A.~D.,  Shafter A.~W.,    Mattei J.~A.,
  1993, PASP, 105, 69

\bibitem[\protect\citeauthoryear{Patterson, Jablonski, Koen, O'Donoghue \&
  Skillman}{Patterson et~al.}{1995}]{pat95v1159ori}
Patterson J.,  Jablonski F.,  Koen C.,  O'Donoghue D.,    Skillman D.~R.,
  1995, PASP, 107, 1183

\bibitem[\protect\citeauthoryear{Patterson, McGraw, Coleman \&
  Africano}{Patterson et~al.}{1981}]{pat81wzsge}
Patterson J.,  McGraw J.~T.,  Coleman L.,    Africano J.~L.,  1981, ApJ, 248,
  1067

\bibitem[\protect\citeauthoryear{Rappaport, Joss \& Verbunt}{Rappaport
  et~al.}{1983}]{rap83CVevolution}
Rappaport S.,  Joss P.~C.,    Verbunt F.,  1983, ApJ, 275, 713

\bibitem[\protect\citeauthoryear{Rappaport, Joss \& Webbink}{Rappaport
  et~al.}{1982}]{rap82CVevolution}
Rappaport S.,  Joss P.~C.,    Webbink R.~F.,  1982, ApJ, 254, 616

\bibitem[\protect\citeauthoryear{Robertson, Honeycutt \& Turner}{Robertson
  et~al.}{1995}]{rob95eruma}
Robertson J.~W.,  Honeycutt R.~K.,    Turner G.~W.,  1995, PASP, 107, 443

\bibitem[\protect\citeauthoryear{Semeniuk, Olech, Kwast \& Nalezyty}{Semeniuk
  et~al.}{1997}]{sem97swuma}
Semeniuk I.,  Olech A.,  Kwast T.,    Nalezyty M.,  1997, Acta Astron.,
  47, 201

\bibitem[\protect\citeauthoryear{Shara, Livio, Moffat \& Orio}{Shara
  et~al.}{1986}]{Hibernation}
Shara M.~M.,  Livio M.,  Moffat A. F.~J.,    Orio M.,  1986, ApJ, 311, 163

\bibitem[\protect\citeauthoryear{Stellingwerf}{Stellingwerf}{1978}]{PDM}
Stellingwerf R.~F.,  1978, ApJ, 224, 953

\bibitem[\protect\citeauthoryear{Udalski}{Udalski}{1990}]{uda90suuma}
Udalski A.,  1990, AJ, 100, 226

\bibitem[\protect\citeauthoryear{Uemura, Kato, Pavlenko, Baklanov \&
  Pietz}{Uemura et~al.}{2001}]{uem01v725aql}
Uemura M.,  Kato T.,  Pavlenko E.,  Baklanov A.,    Pietz J.,  2001, PASJ, 53,
  539

\bibitem[\protect\citeauthoryear{van~der Woerd, van~der Klis, van Paradijs,
  Beuermann \& Motch}{van~der Woerd et~al.}{1988}]{vanderwoe88lateSH}
van~der Woerd H.,  van~der Klis M.,  van Paradijs J.,  Beuermann K.,    Motch
  C.,  1988, ApJ, 330, 911

\bibitem[\protect\citeauthoryear{Vogt}{Vogt}{1980}]{vog80suumastars}
Vogt N.,  1980, A\&A, 88, 66

\bibitem[\protect\citeauthoryear{Vogt}{Vogt}{1983}]{vog83lateSH}
Vogt N.,  1983, A\&A, 118, 95

\bibitem[\protect\citeauthoryear{Vogt \& Bateson}{Vogt \&
  Bateson}{1982}]{vog82atlas}
Vogt N.,  Bateson F.~M.,  1982, A\&AS, 48, 383

\bibitem[\protect\citeauthoryear{Warner}{Warner}{1985}]{war85suuma}
Warner B.,  1985, in Eggelton P.~P.,  Pringle J.~E.,  eds, Interacting Binaries
  (Dordrecht: D. Reidel Publishing Company), p.~367

\bibitem[\protect\citeauthoryear{Warner}{Warner}{1995}]{war95suuma}
Warner B.,  1995, Ap\&SS, 226, 187

\end{thebibliography}
\end{document}